\documentclass[9pt,conference,a4paper]{IEEEtran}
\usepackage{cite}

\usepackage{aas_macros}

\ifCLASSINFOpdf
  \usepackage[pdftex]{graphicx}
\else
\fi

\begin{document}

\title{PURIFYing real radio interferometric observations}

\author{%
\IEEEauthorblockN{
Luke Pratley\IEEEauthorrefmark{1}, 
Jason D.~McEwen\IEEEauthorrefmark{1}, 
Mayeul d'Avezac\IEEEauthorrefmark{2}, 
Rafael E.~Carrillo\IEEEauthorrefmark{3},
Alexandru Onose\IEEEauthorrefmark{4},
Yves Wiaux\IEEEauthorrefmark{4}
}
\IEEEauthorblockA{\IEEEauthorrefmark{1} 
Research Software Development Group, Research IT Services
University College London (UCL),
London WC1E 6BT
UK}
\IEEEauthorblockA{\IEEEauthorrefmark{2} 
Mullard Space Science Laboratory (MSSL), 
University College London (UCL),
Holmbury St Mary, Surrey RH5 6NT,
UK}
\IEEEauthorblockA{\IEEEauthorrefmark{3}
Signal Processing Laboratory (LTS5), 
Ecole Polytechnique F\'ed\'erale de Lausanne (EPFL),
Lausanne CH-1015, 
Switzerland
}
\IEEEauthorblockA{\IEEEauthorrefmark{4}
Institute of Sensors, Signals, and Systems, 
Heriot-Watt University, 
Edinburgh EH14 4AS, 
UK}
}
\maketitle

\begin{abstract}
Next-generation radio interferometers, such as the Square Kilometre Array (SKA), will revolutionise our understanding of the universe through their unprecedented sensitivity and resolution. However, standard methods in radio interferometry produce reconstructed \mbox{inter\-ferometric} images that are limited in quality and they are not scalable for big data. In this work we apply and evaluate alternative interferometric reconstruction methods that make use of state-of-the-art sparse image reconstruction algorithms motivated by compressive sensing, which have been implemented in the PURIFY software package. In particular, we implement and apply the proximal alternating direction method of multipliers (P-ADMM) algorithm presented in a recent article. We apply PURIFY to real interferometric observations. For all observations PURIFY outperforms the standard CLEAN, where in some cases PURIFY provides an improvement in dynamic range by over an order of magnitude. The latest version of PURIFY, which includes the developments presented in this work, is made publicly available.
\end{abstract}

\section{Introduction}
Radio interferometry allows imaging of the radio universe at higher resolution and sensitivity than possible with a single radio telescope. Image reconstruction methods are needed to reconstruct the true sky brightness distribution from the raw data acquired by the telescope, which amounts to solving an ill-posed inverse problem. Traditional methods, which are mostly variations of the H\"{o}gbom CLEAN algorithm \cite{hog74}, do not exploit modern state-of-the-art image reconstruction techniques.

Next-generation radio interferometers, such as the Square Kilometer Array (SKA; \cite{dew13}), must meet the challenge of processing and imaging extremely large volumes of data. These experiments have ambitious, high-profile science goals, including detecting the Epoch of Re-ionisation (EoR) \cite{koo15}. If these science goals are to be realised, state of the art methods in image reconstruction are needed to process big data and to reconstruct images with high fidelity.

\begin{figure}
\center
    \includegraphics[width=1\linewidth]{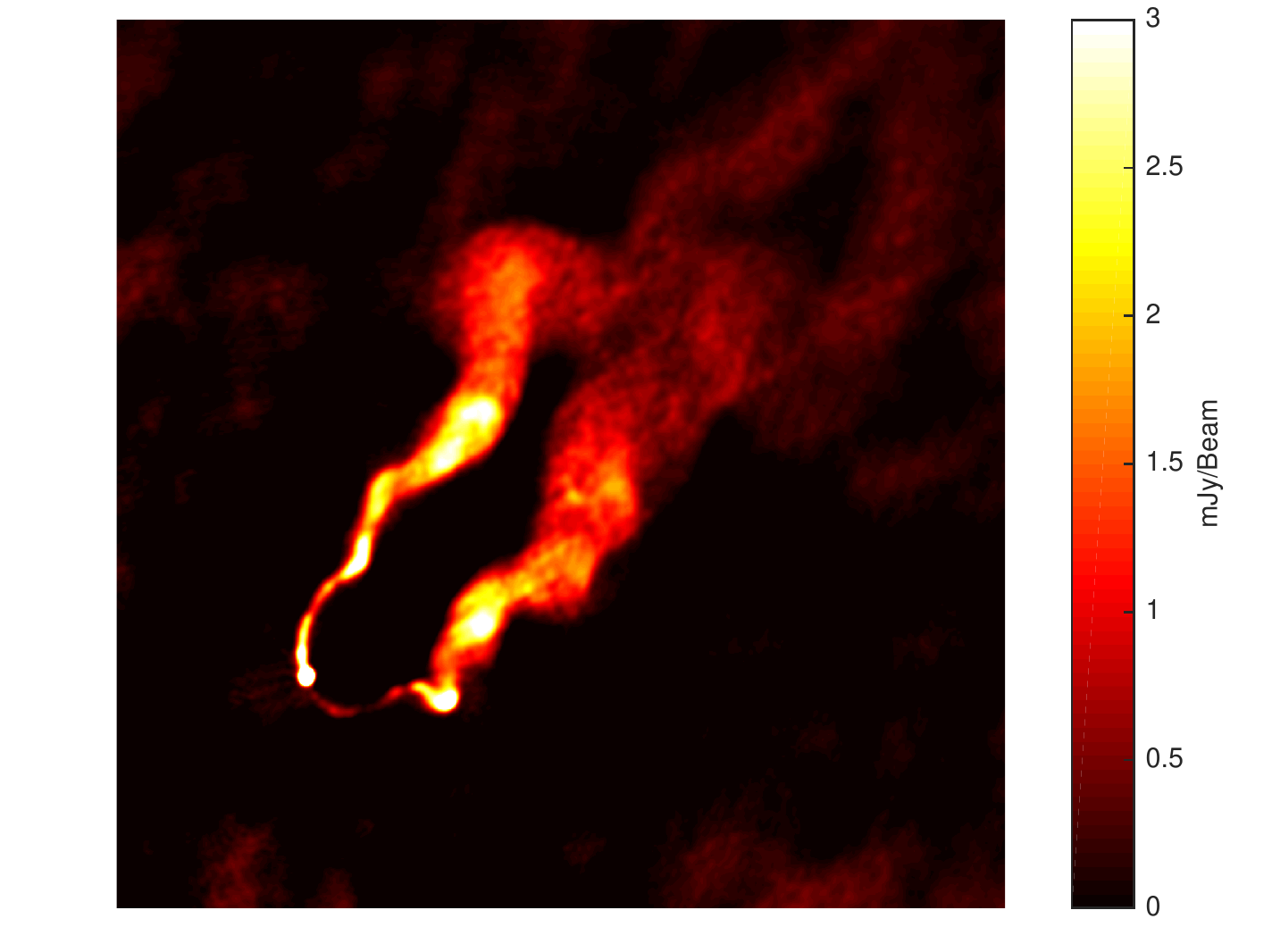}
    \includegraphics[width=1\linewidth]{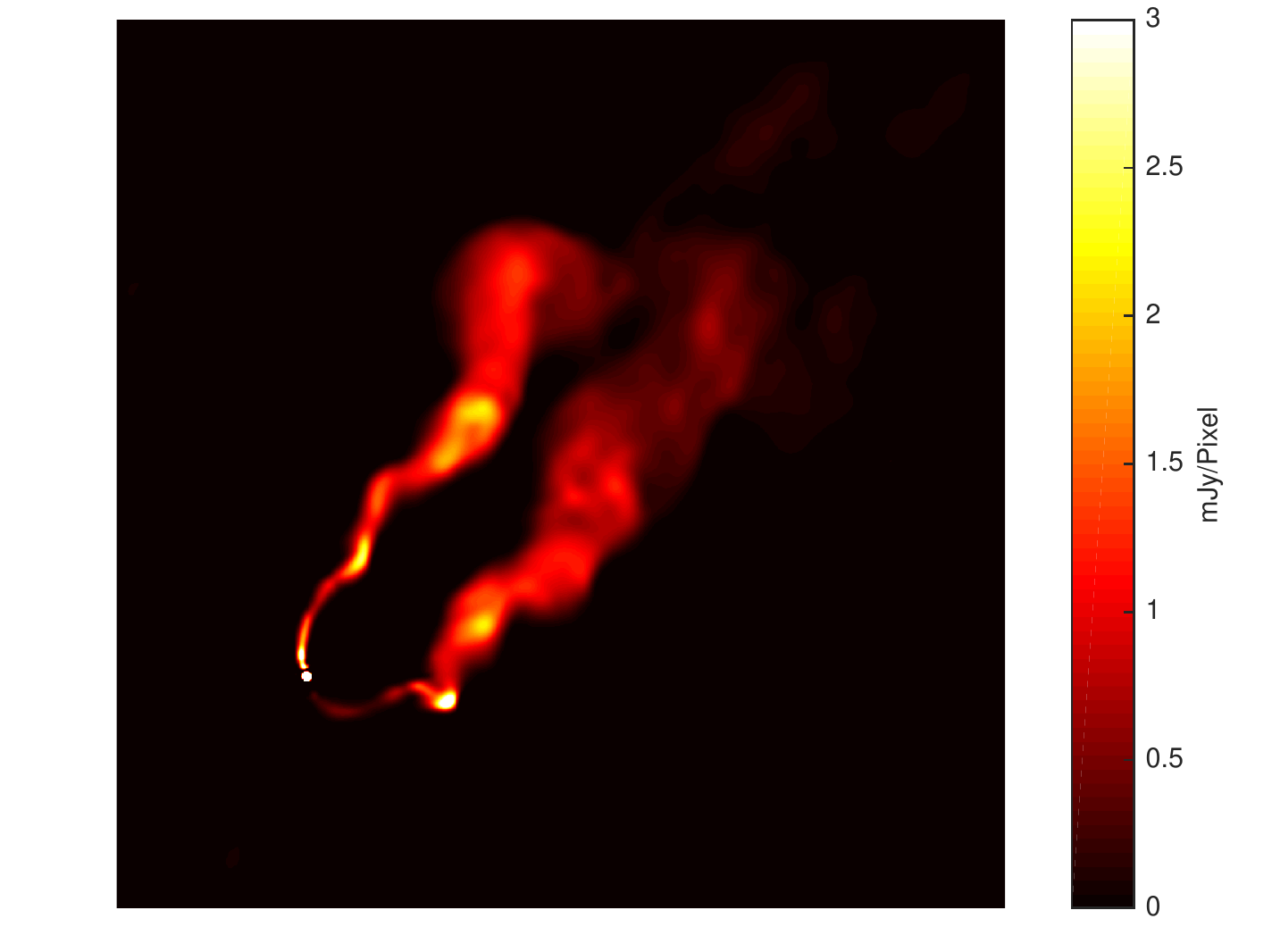}
		\caption{The top and bottom show CLEAN and PURIFY reconstructions of the radio galaxy 3C129 respectively. The PURIFY recosntruction shows less contamination with higher dynamic range than the CLEAN reconstruction. Additionally, the PURIFY reconstruction does not require post processing to create a restored image. The details of these reconstructions can be found in \cite{pra16a}.}
		\label{fig:3C129}
\end{figure}

In \cite{pra16a} we implement the P-ADMM algorithm developed by \cite{ono16} in the PURIFY software package, which has been entirely redesigned and re-implemented in C++, and apply it to observational data from the VLA and the ATCA. The previous version of \mbox{PURIFY} supported only simple models of the measurement operator modelling the telescope.  PURIFY now supports a wider range of more accurate convolutional interpolation kernels (for gridding and degridding). We found that the Kaiser-Bessle kernel performs as well as the prolate spheroidal wave funtions. Additionally, PURIFY provides higher dynamic range images than CLEAN on real observations, sometimes by an order of magnitude improvement.
Figure ~\ref{fig:3C129} shows an example of a CLEAN and PURIFY reconstruction of the radio galaxy 3C129. The PURIFY reconstructions produce higher dynamic range without the need for post processing to create a restored image.

\bibliographystyle{IEEEtran}
\bibliography{ref}

\end{document}